\documentclass[11pt,a4]{article}
\usepackage{amssymb,amsthm,url,paracol}
\usepackage{graphicx,hyperref,url}
\usepackage{longtable}
\usepackage[english]{babel}
\usepackage{authblk}

\hypersetup{
    colorlinks = true,
    linkcolor = blue,
    anchorcolor = blue,
    citecolor = blue,
    filecolor = blue,
    urlcolor = blue
    }
    
\renewcommand*{\Affilfont}{\footnotesize}

\title{A New Sample of Gamma-Ray Emitting Jetted Active Galactic~Nuclei\footnote{Accepted for publication on {\em Universe} journal, Special Issue on \href{https://www.mdpi.com/journal/universe/special_issues/BHRJ}{\em Black Holes and Relativistic Jets}, edited by I. Dutan and N. R. MacDonald. This is only the main text. The tables A1 and A2 are available on the \href{https://www.mdpi.com/2218-1997/8/11/587}{journal web site} (open access).}}

\author[1]{Luigi Foschini\footnote{Email: {\tt luigi.foschini@inaf.it}; Tel.: +39-02-72320-458.}}
\author[2]{Matthew L. Lister}
\author[3]{Heinz Andernach}
\author[4]{Stefano Ciroi}
\author[5]{Paola Marziani}
\author[6]{Sonia Ant\'on}
\author[7]{Marco Berton}
\author[4]{Elena Dalla~Bont\`a}
\author[8]{Emilia J\"arvel\"a}
\author[9]{Maria J.~M. March\~{a}}
\author[1]{Patrizia Romano}
\author[10]{Merja Tornikoski}
\author[1]{Stefano Vercellone}
\author[4]{Amelia Vietri}

\affil[1]{\Affilfont Osservatorio Astronomico di Brera, Istituto Nazionale di Astrofisica (INAF), 23807 Merate, Italy.}
\affil[2]{Department of Physics and Astronomy, Purdue University, West Lafayette, IN 47907, USA.}
\affil[3]{Departamento de Astronom\'ia, Universidad Guanajuato, Callej\'on de Jalisco s/n, Guanajuato 36023, GTO, Mexico.}
\affil[4]{Dipartimento di Fisica e Astronomia, Universit\`a di Padova, 35122 Padova, Italy.}
\affil[5]{Osservatorio Astronomico di Padova, Istituto Nazionale di Astrofisica (INAF), 35122 Padova, Italy.}
\affil[6]{Centro de F\'isica da UC, Departamento de F\'{i}sica, Universidade de Coimbra, 3004-516 Coimbra, Portugal.}
\affil[7]{European Southern Observatory (ESO), Santiago de Chile 19001, Chile.}
\affil[8]{European Space Astronomy Centre (ESAC), European Space Agency (ESA), 28692 Villanueva de la Ca\~{n}ada, Spain.}
\affil[9]{Physics and Astronomy Department, University College London, London WC1E 6BT, UK.}
\affil[10]{Mets\"{a}hovi Radio Observatory, Aalto University, 02540 Kylm\"{a}l\"{a}, Finland.}

\begin{document}
\maketitle

\begin{abstract}
We considered the fourth catalog of gamma-ray point sources produced by the \emph{Fermi} Large Area Telescope (LAT) and selected only jetted active galactic nuclei (AGN) or sources with no specific classification, but with a low-frequency counterpart. Our final list is composed of 2980 gamma-ray point sources. We then searched for optical spectra in all the available literature and publicly available databases, to measure redshifts and to confirm or change the original LAT classification. Our final list of gamma-ray emitting jetted AGN is composed of BL Lac Objects (40\%), flat-spectrum radio quasars (23\%), misaligned AGN (2.8\%), narrow-line Seyfert 1, Seyfert, and low-ionization nuclear emission-line region galaxies (1.9\%). We also found a significant number of objects changing from one type to another, and vice versa (changing-look AGN, 1.1\%). About 30\% of gamma-ray sources still have an ambiguous classification or lack one altogether.
\end{abstract}

\section{Introduction}
The current paradigm of jetted active galactic nuclei (AGN) is mostly rooted in the seminal works by Rees, Schmidt, Blandford, Fanaroff, Riley, Orr, Browne, Barthel, Urry, Padovani, Ghisellini, just to cite a few \cite{REES,SCHMIDT,FR,BLANDREES,ORRBROWNE,BARTHEL,URRYPAD,GHISELLINI98,PADOVANI}. Jetted AGN are basically divided into two main classes depending on the jet viewing angle (aligned with the Earth or not), which in turn are divided into two subclasses depending on the accretion rate. Flat-spectrum radio quasars (FSRQ) and BL Lac Objects have a small jet viewing angle, but the former have disks accreting at high rate, while the latter have weak and inefficient disks. They form the so-called blazar sequence, with FSRQs on one side, emitting high jet power, and BL Lac Objects on the opposite side, with low jet power. Misaligned AGN are commonly called radio galaxies, and are also divided according to the accretion rate into High-Excitation Radio Galaxies (HERG) and Low-Excitation Radio Galaxies (LERG). All these objects are powered by central supermassive black holes ($M\gtrsim 10^{8}M_{\odot}$) hosted in giant elliptical galaxies (see \cite{BLANDFORD} for a recent review). 

However, the discovery of powerful relativistic jets from Narrow-Line Seyfert 1 galaxies (NLS1s) proved that the zoo of jetted AGN is more variegated than previously thought (see, for example, \cite{KOMOSSA18,FOSCHINI20} for recent reviews). Although, NLS1s have been proven to be the low-luminosity tail of the FSRQs distribution \cite{BERTON16}, the relatively small mass of their central black hole and the high accretion rate implied that the blazar sequence no longer stands \cite{FOSCHINI17}. Therefore, it is important to keep the NLS1s classification separated from that of FSRQs, to avoid losing important physical information and implications, such as the second branch in the Jet-Disk plane (JD-plane), the branch of small-mass/high-accretion AGN \cite{FOSCHINI17}.

Today, understanding the impact of NLS1s on the population of gamma-ray sources is hampered by the small number of known objects of this type ($\sim 20$ \cite{ROMANO18}). In addition, recent studies on large samples are done by using computer-based procedures designed according to the old paradigm, which implies that this new class of objects is not recognized. Therefore, in order to have a large sample of gamma-ray emitting jetted AGN with updated and reliable optical classification and spectroscopic redshift, we performed the reclassification of the gamma-ray sources in the {\em Fermi} Large Area Telescope (LAT) (4FGL-DR2,~\cite{4FGL}, 4LAC,~\cite{4LAC}) with extragalactic or unclassified counterparts (with the exclusion of starburst and normal galaxies), and outside the Galactic plane ($|b|>10^{\circ}$). We collected 2980 gamma-ray point sources\footnote{The original sample from 4FGL-DR2 consisted of 2982 point sources \cite{FOSCHINI21}, but we updated it to 2980 when the DR3 was released, because J$1242.4-2948$ has no longer a counterpart and J$2055.8+1545$ is now associated with a millisecond pulsar. We did not include new sources added in the DR3. Hereinafter reference is always made to DR2, unless otherwise specified.}. Then, for each source we searched for redshift measurements and optical spectra in the literature and data through the following public databases: 

\begin{itemize}
\item Set of Identifications, Measurements and Bibliography for Astronomical Data (SIMBAD\footnote{\url{http://simbad.u-strasbg.fr/simbad/} (accessed on 30 August 2022).});
\item NASA/IPAC Extragalactic Database (NED\footnote{\url{http://ned.ipac.caltech.edu/} (accessed on 30 August 2022).});
\item SAO/NASA Astrophysics Data System (ADS\footnote{\url{https://ui.adsabs.harvard.edu/} (accessed on 30 August 2022).});
\item Sloan Digital Sky Survey (SDSS DR16\footnote{\url{http://skyserver.sdss.org/DR16/en/home.aspx} (accessed on 30 August 2022).});
\item Large Sky Area Multi-Object Fiber Spectroscopic Telescope (LAMOST DR6V2\footnote{\url{http://dr6.lamost.org/v2/} (accessed on 30 August 2022).}). 
\end{itemize}

Preliminary results of this work (right ascension $0^{\rm h}$--$12^{\rm h}$, J2000) have been published in 2021 (\cite{FOSCHINI21}, Paper I hereafter), and we refer to that paper for more details on the adopted procedures and explanations of the new classes of AGN. 

\section{Classification and Redshift}
\begin{figure}[h]
\centering
\includegraphics[scale=0.2]{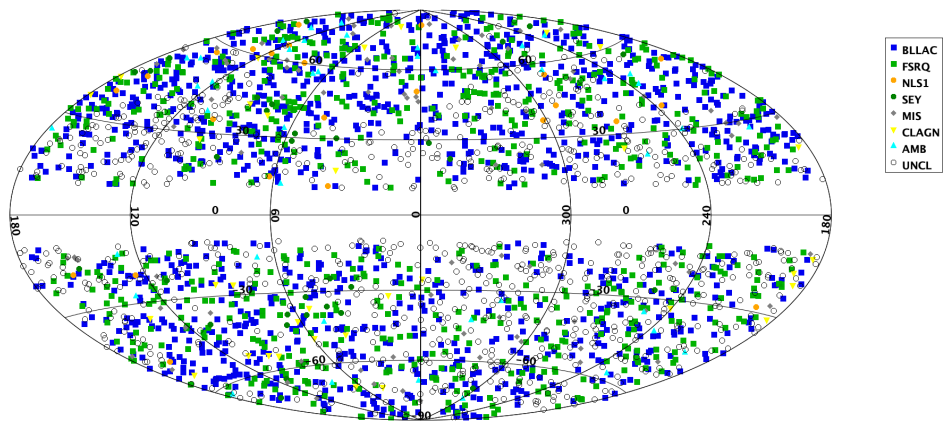}
\caption{Distribution of the present list of gamma-ray sources in the sky (Galactic coordinates, Aitoff projection) according to the new classification. \label{display}}
\end{figure}

The full list of sources with their new classification is available in Appendix~A\footnote{The full tables A1 and A2 are available in the \href{https://www.mdpi.com/2218-1997/8/11/587}{open-access paper}.}. Table~\ref{distrib} summarizes the statistics of gamma-ray emitting jetted AGN after our reclassification, the fraction of sources with spectroscopic redshift, and the statistics from the original 4FGL-DR2 catalog. The sky distribution (Galactic coordinates, Aitoff projection) is shown in Figure~\ref{display}. It is worth noting that there are some differences of classes between our classification and that of 4FGL, as explained in the notes of Table~\ref{distrib}. We refer to Paper I for more details. 

\begin{figure}[h]
\centering
\includegraphics[scale=0.2]{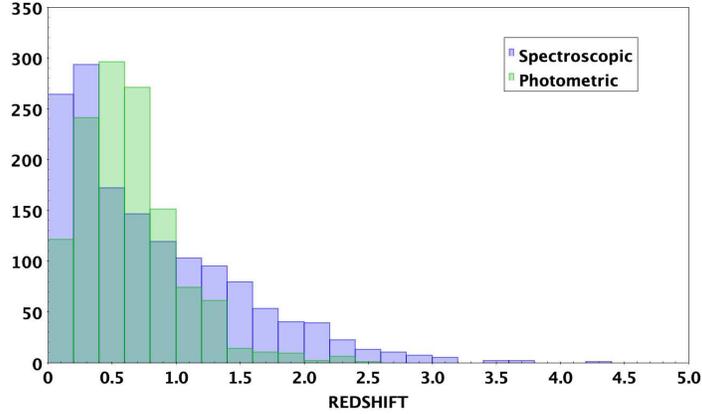}
\caption{Distribution of the redshifts (spectroscopic and photometric). \label{distrz}}
\end{figure}

We also searched for photometric redshifts $z_{\rm p}$ from a variety of catalogs and found at least one value for 2631 sources (88\%, see Table~A2). The complete (spectroscopic plus photometric) redshift distribution of sources is displayed in Figure~\ref{distrz}. This information must be considered with care, because we noted some discrepancies between the coordinates of the counterparts given in the 4FGL and those available in radio databases. We generally considered valid the 4FGL coordinates, although we point out some cases of significant offsets (see also Section~\ref{cave}).

\begin{table}[h]
\caption{Distribution of gamma-ray emitting jetted AGN according to the present work and comparison with the original 4FGL subsample. Columns: (1) Classification according to our criteria; (2) number N of sources of the corresponding class; (3) percentage of sources with spectroscopic redshift $z$; (4) number of sources with the same (or similar) classification in the 4FGL. The notes at the end of the table explain the differences between the present classification criteria and those of the 4FGL.  \label{distrib}}
\begin{tabular}{p{6cm}p{1.5cm}p{2cm}p{2cm}}
\hline
\textbf{Classification}				& \textbf{N} & \textbf{{\em z}} & \textbf{4FGL}\\
\hline
BL Lac Object (BLLAC)				& 1207 & 47.2\% 						& 1204\\
Flat-spectrum radio quasar (FSRQ) 	& 695 & 99.7\%			& 703\\
Misaligned AGN (MIS $^1$) 			& 85 & 96.5\%         & 45\\
Narrow-Line Seyfert 1 galaxy (NLS1) & 24 & 100\% & 9\\
Seyfert galaxy (SEY $^2$) 			& 32 & 100\% & 0\\ 
Ambiguous (AMB) 					& 42 & 69\% & - \\
Changing-look AGN (CLAGN) 			& 34 & 100\% & -\\
Unclassified (UNCL $^3$) 			& 861 & 0.1\% & 1009 \\
\hline
Total & 2980 & 49\% $^4$ & 2980 $^5$\\ 
\hline
\end{tabular}\\
\footnotesize
$^1$ The MIS class includes the 4FGL classes RDG/rdg, SSRQ/ssrq, and CSS/css. $^2$ Seyfert class includes Seyfert 1, 2, intermediate, and LINERS. $^3$ The UNCL class includes the 4FGL classes BCU/bcu, and UNK/unk. $^4$ Photometric redshift is also available for another 43\% of sources. Therefore, 93\% of sources do have a redshift value. $^5$ 4FGL contains also 10 sources classified as non-blazar active galaxy (AGN/agn), which were reclassified as BLLAC (1), MIS (3), CLAGN (3), UNCL (3).\normalsize
\end{table}

\begin{figure}[h]
\centering
\includegraphics[scale=0.2]{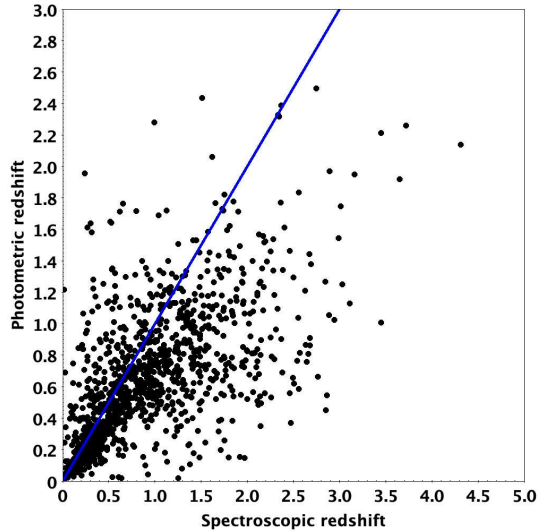}
\caption{Comparison of photometric redshift with spectroscopic one, for those sources having both values. The blue line shows the function $y=x$. \label{photospec}}
\end{figure}

\begin{figure}[h]
\centering
\includegraphics[scale=0.2]{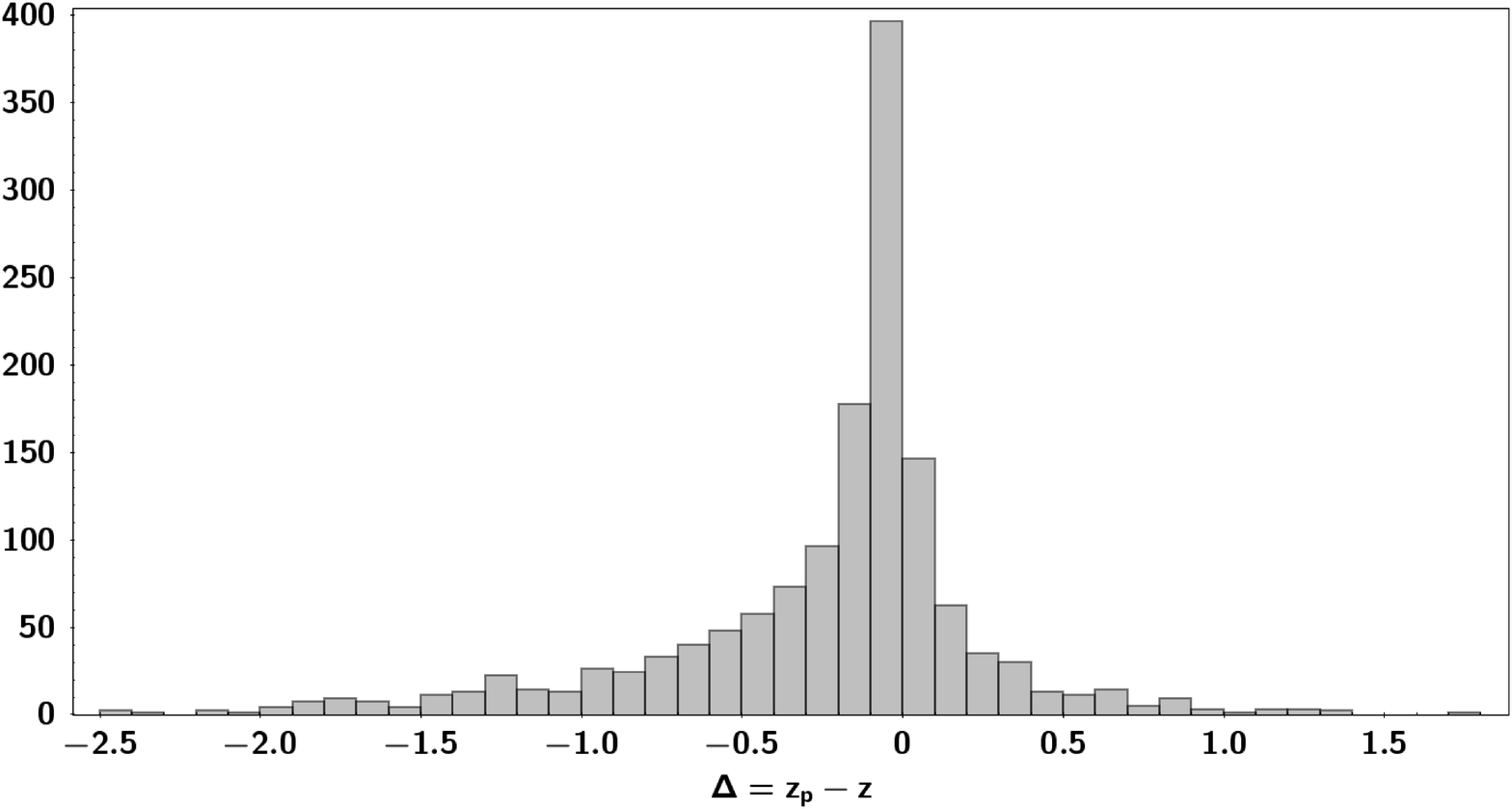}
\caption{Distribution of the difference $\Delta=z_{\rm p}-z$ between the photometric and spectroscopic redshifts.  \label{histored}}
\end{figure}

Although in Table~A1 we include photometric redshifts only for the sources without a spectroscopic one (42\% of the total sample), the availability of both measurements for a large sample of sources allowed us to give a rough estimation of the reliability of $z_{\rm p}$. As shown in Figure~\ref{photospec}, there is some linear relationship between the two measurements for values smaller than one, but, for greater values, $z_{\rm p}$ tends to be underestimated with respect to the spectroscopic measurements: while $z_{\rm p}\lesssim 2.5$, the spectroscopic redshift reached values up to $\sim$4.3. Figure~\ref{histored} displays the distribution of the difference $\Delta=z_{\rm p}-z$, between photometric and spectroscopic redshifts: in most of cases, the photometric redshift is underestimated with respect to the spectroscopic one by $\Delta\lesssim -0.1$. 

It is also worth reporting the breakdown of our new classes, according to more common notations. For example, the MIS class contains 85 sources divided into: 8 Fanaroff-Riley type 0 (FR0), 35 FRI, 18 FRII, 7 compact steep-spectrum sources (CSS), and 1 steep-spectrum radio quasar (SSRQ). For the remaining 16 sources, we did not find publications with specific identification according to the above cited subclasses. 

The AMB class contains a very heterogeneous set of sources, because the ambiguity can be due to different reasons:

\begin{itemize}
\item the difficulty to have a clear measure of the viewing angle, and so to distinguish between beamed and unbeamed jets (18 cases);
\item different values of spectroscopic redshift, but no spectra published, making it impossible to choose the more reliable value (9);
\item only a value of redshift without published spectra or any information about lines (3);
\item Seyfert vs. NLS1, when no measurement of the full width half maximum (FWHM) of the H$\beta$ emission line is available (5);
\item issues in the counterpart coordinates, see Section~\ref{cave} (6); 
\item the possibility that the counterpart might be a Galactic source (1).
\end{itemize}

Further studies might solve these issues and change the number of objects in one class or another.

Most of CLAGN are beamed jetted AGN (30), transitioning from a featureless continuum to a line-dominated spectrum or vice versa, or displaying a change of the spectral energy distribution (e.g., J$2345.2-1555$ alias PMN~J$2345-1555$ \cite{GHISELLINI}). Three cases are misaligned AGN (J$0014.2+0854=$~MS~$0011.7+0837$; J$0522.9-3628=$~PKS~$0521-36$; J$0910.0+4257=$~3C~$216$), and one curious case (J$2334.9-2346$ alias PKS~$2331-240$) refers to a change of the jet viewing angle, from a misaligned (MIS) to an aligned source (SEY)~\cite{HERNANDEZ}. It is difficult to establish the real impact of CLAGN on the overall classification and population statistics, because most of the sources in the present sample do have only one optical spectrum. However, this is a very important point: the apparent classification is time-dependent and it would be desirable to move to more physics-based classifications.

\section{Caveats}\label{cave}
We have already pointed out some sources classified as AMB, because there were problems in the coordinates of the counterpart. For example, the 4FGL coordinates of the counterpart of J$0438.7-3441$ differ by $\sim 36'$ from those of the gamma-ray centroid and located far outside the 95\% error ellipse ($3.6'\times 3.1'$). There is only one radio source inside the error ellipse of the gamma-ray source, so that we changed the 4FGL counterpart with this radio source and set the classification as AMB, because it needs more study to be confirmed or rejected. 

It is worth citing another complex example: J$2127.6-5959$ is associated with NGC~$7059$, a nearby starforming spiral galaxy ($z=0.00578$). However, the 4FGL coordinates of the counterpart are not consistent with the center of the galaxy (difference $\sim 1'$). These coordinates are consistent with a \emph{ROSAT} source, 1RXS~J$212728.9-600049$ (error radius~$\sim 15''$, therefore not consistent with the galaxy center\footnote{Although NED considers 1RXS~J$212728.9-600049$ as the X-ray counterpart of NGC~$7059$, but, as we have shown, this is not the case.}). \emph{Swift} follow-up of the gamma-ray source suggested a slightly different counterpart Swift J$212729.3-600102$ (distant $\sim 13''$ from the \emph{ROSAT} source), although consistent within the position errors with the \emph{ROSAT} source~\cite{KAUR19,KERBY21}. However, at radio frequencies, there are two counterparts observed at
$944$~MHz with the Australian SKA Pathfinder (ASKAP) Evolutionary Map of the Universe (EMU, \cite{NORRIS21}) Pilot Survey: one is consistent with the centroid of 1RXS~J$212728.9-600049$ ($\sim 2$~mJy flux density and deconvolved size of $\sim 27'' \times 10'$); the other is consistent with the centroid of Swift~J$212729.3-600102$ ($\sim 2.9$~mJy flux density and is only barely resolved, $\sim 9'' \times 7''$, J. Marvil, NRAO, priv. comm.). Therefore, more detailed studies are needed to assess the real counterpart of the gamma-ray source 4FGL~J$2127.6-5959$. 

These examples show that sometimes we found inconsistencies between the name of the associated counterpart and its coordinates, but we decided to keep as reference the coordinates, and wrote potential issues in the notes. We also noted some minor differences (at arcsecond level) between the 4FGL coordinates and the values reported in radio catalogs, but again we kept 4FGL as reference, because this type of investigation is beyond our aims. However, we note that this discrepancy may affect the photometric redshifts reported in Table~A2.

Another caveat refers to the classification. As previously stated, we made this reclassification almost completely according to the published information. This does not imply that it is carved into the stone. Particularly, large data sets cannot be analyzed directly by human being\footnote{The reclassification of the present sample of 2980 sources on a case-by-case basis required more than three years of one person's almost full time.} and require computer-aided procedures, which in turn---being prepared having in mind certain quantities and characteristics---can easily miss interlopers and outliers. Therefore, computer-aided analyses must be always verified, particularly if the spectrum displays strong noise or distortion of the line profiles. 

\begin{figure}[h]
\centering
\includegraphics[scale=0.2]{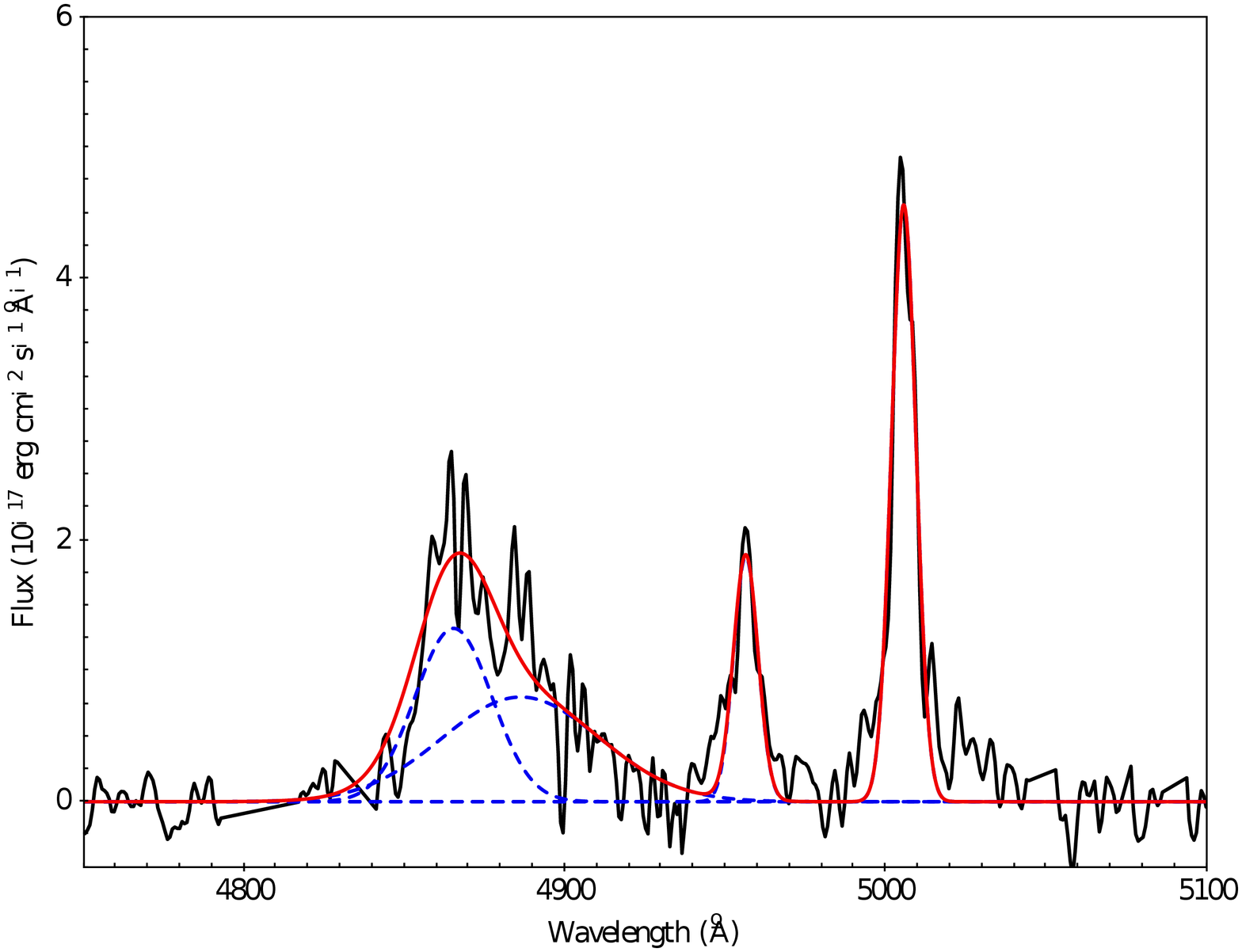}
\includegraphics[scale=0.2]{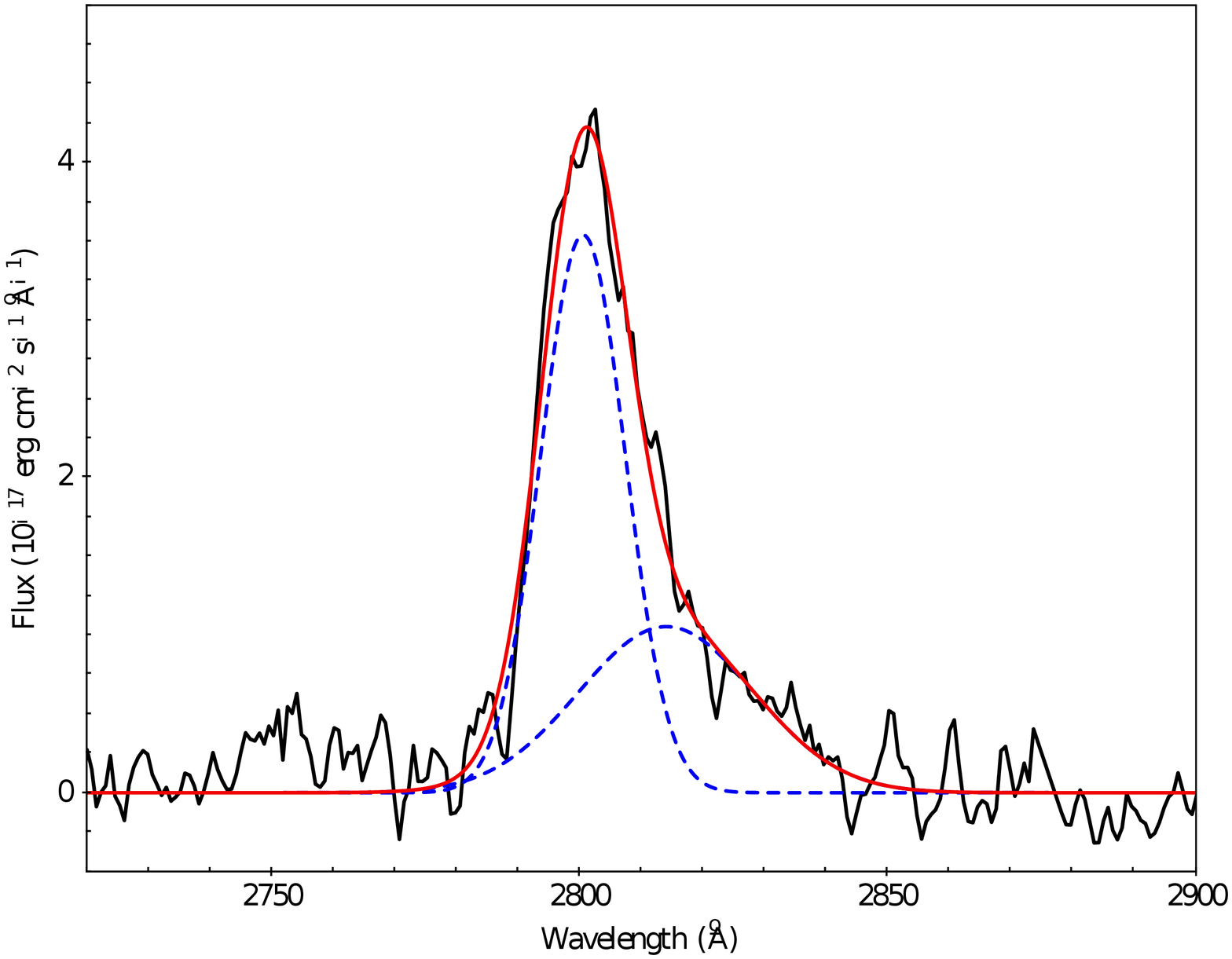}
\caption{Reanalysis of SDSS spectrum of PKS~$1441+25$: ({\em left panel}) H$\beta$ and [OIII] complex; ({\em right panel}) MgII. Both H$\beta$ and MgII display a clear red wing, and were fitted with a narrower component plus a broader red one (blue dashed lines). The SDSS spectrum is represented by a black continuous line, while the individual components are depicted with dashed blue lines. The model sum of the different components is a red continuous line. \label{pks1441}}
\end{figure}

The case of J$1443.9+2501$ (PKS~$1441+25$, $z=0.940$) is exemplary: Shaw et~al.~\cite{SHAW12} measured FWHM(H$\beta$)$~=~1600\pm400$~km/s from a very noisy spectrum, with H$\beta$ barely visible and flooded in a strong background (the spectrum is available only in the online version of Shaw's work\footnote{\url{https://iopscience.iop.org/article/10.1088/0004-637X/748/1/49} (accessed on 30 August 2022).}). 
The SDSS spectrum\footnote{\url{http://skyserver.sdss.org/DR16//en/tools/explore/summary.aspx?ra=220.9871&dec=25.029} (accessed on 30 August 2022). 
} is a bit better, and clearly shows the H$\beta$-[OIII] complex. Rakshit et al. \cite{RAKSHIT20} performed the measurement of the spectral properties of a large sample of AGN in the SDSS DR14. For PKS~$1441+25$, they measured FWHM(H$\beta$)$~=~1962\pm433$~km/s. Were these measurements correct, this AGN should be classified as NLS1, an unexpected and great result, because PKS~$1441+25$ was detected in 2015 at Very High Energies (VHE) by the MAGIC telescope \cite{MAGIC15}. Therefore, given the importance of the possible result, we reanalyzed the publicly available SDSS spectrum. The H$\beta$ profile is significantly distorted, with an apparent red wing, as like as the MgII. The line shape was decomposed into a narrow blue component and a broad red one. We measured FWHM(H$\beta_{\rm n,blue}$)~$\sim 1700$~km/s for the former, and FWHM(H$\beta_{\rm b,red}$)~$\sim 3500$~km/s for the latter (see Figure~\ref{pks1441}), thus rejecting the NLS1 classification and to confirm the FSRQ one. These profiles can be fit with a relativistic accretion disk model oriented almost face-on, with $R_{\rm in}=250r_{\rm g}$ ($r_{\rm g}$ is the gravitational radius), $R_{\rm out}=3000r_{\rm g}$, $5^{\circ}$ inclination, emissivity exponent $a=2.5$ (continuum power law $\propto R^{-a}$), and local dispersion of H$\beta$ $\sigma/\nu_{0}=2.8\times 10^{-3}$ \cite{CHEN89}. These parameters are consistent with their aligned classification, and with the current interpretation of the quasar main sequence \cite{SHENHO,PANDA}: for low FeII emission the line width is governed mainly by orientation (see Figure 3 of \cite{MARZIANI22}). Similar cases of distorted emission-line profiles has been detected in many FSRQs and are likely due to low-ionization outflows \cite{MARZIANI13} or gravitational redshift \cite{PUNSLY20}. It is also worth noting that NLS1s do generally have Lorentzian profiles and the significant red wing apparent in the SDSS spectrum was another point against the NLS1 classification. To estimate the mass of the central black hole, it is better to use the [OIII], given the significant distortion of the H$\beta$ line profile. We measured a FWHM([OIII])~$\sim 500$~km/s and, by applying the $M-\sigma_{*}$ relationship \cite{HO14} (see also Equation~(6) in \cite{BERTON15}), we can estimate $M\sim 4\times 10^{8}M_{\odot}$, which is typical for FSRQs.

We already started a parallel follow-up program to reanalyze the publicly available optical spectra for the NLS1 and SEY classes, and to ask for new high-quality observations. The new data, when available, will be reported elsewhere. Here we just want to remind that the present results have to be taken {\em cum grano salis}.

Last, but not least, works like the present one never end. New observations can improve or reject the current classification and the gamma-ray sky is still an effervescent research field, so that new papers are published at a non-negligible rate. The present work includes information published until 30 August 2022. 

\section{Comparison with CGRO/EGRET}
The present reclassification resulted in 24 NLS1s and 32 Seyfert/LINERs candidates or confirmed ones. Therefore, it is rather obvious to wonder if Seyfert-type AGN could have been detected by the EGRET instrument onboard the \emph{Compton Gamma-Ray Observatory}. Although it is expected that the jet power of highly-accreting small-mass black holes hosted by NLS1s scales with $M^{17/12}$ \cite{HEINZ}, it is also known that strong gamma-ray outbursts have been observed from NLS1 by {\em Fermi}/LAT, with fluxes exceeding $\sim$$10^{-6}$~ph~cm$^{-2}$~s$^{-1}$ at energies greater than $100$~MeV (e.g., \cite{ABDO09,FOSCHINI11,PALIYA14,PALIYA16}). These values are within the capabilities of {\em CGRO}/EGRET. Therefore, we cross-matched our reclassified list of gamma-ray sources with the Third EGRET Catalog \cite{EGRET} by using an error circle of one degree. We found 100~matches, subdivided into 54 FSRQ, 30 BLLAC, 2 NLS1, 2 MIS, 1 AMB, 3 CLAGN, and 8~UNCL. 

The two NLS1 are: 

\begin{enumerate}
\item 4FGL~J$0001.5+2113$~=~3EG~J$2359+2041$: in this case, the EGRET source was originally associated with the FSRQ TXS~$2356+196$ ($z=1.07$), while the LAT source, with an improved error circle, has a different counterpart, TXS~$2358+209$ ($z=0.439$). The SDSS spectrum\footnote{\url{http://skyserver.sdss.org/DR16//en/tools/explore/summary.aspx?ra=0.384875&dec=21.226739} (accessed on 30 August 2022).} of the latter clearly displays a H$\beta$ with a Lorentzian profile and the FeII bumps. The analysis by Wu \& Shen \cite{WU22} resulted in FWHM(H$\beta$)$~=~1766\pm316$~km/s and an estimated mass of the central black hole of $\sim 5\times 10^{7}M_{\odot}$. 

\item 4FGL~J$0442.6-0017$~=~3EG~J$0442-0033$: the gamma-ray source is associated in both cases with PKS~$0440-00$ ($z=0.844$), which was classified as AGN \cite{EGRET}, and later as FSRQ \cite{EGRET2}. Shaw \cite{SHAW12} measured FWHM(H$\beta$)~$=~1700\pm1100$~km/s, formally NLS1, but the error is so large to cast significant doubts. A multiwavelength study of this jetted AGN favoring the NLS1 classification has been recently presented by Jessica Luna at the workshop {\em Panchromatic View of the Life-Cycle of AGN} (14--16 September 2022, ESA/ESAC, Spain)\footnote{\url{https://www.cosmos.esa.int/web/life-cycle-of-agn/home} (accessed on 30 August 2022).}. However, an optical spectrum with better S/N is needed to confirm this classification. 
\end{enumerate}

It is also worth noting another case: 4FGL~J$1321.1+2216$~=~3EG~J$1323+2200$, associated with the FSRQ TXS~$1324+224$ ($z=1.4$) by EGRET and to a different counterpart by LAT, TXS~$1318+225$ ($z=0.946$). For the latter, we found three measurements of the FWHM(H$\beta$): $1700\pm 300$~km/s \cite{SHAW12}, $5377\pm 843$~km/s \cite{RAKSHIT20}, and $3725\pm 412$~km/s \cite{WU22}. Therefore, Shaw's measurement suggested it might be a NLS1, but a quick look at the SDSS spectrum\footnote{\url{http://skyserver.sdss.org/DR16//en/tools/explore/summary.aspx?ra=200.2967&dec=22.27} (accessed on 30 August 2022).} shows evident red wings in the profiles of MgII and H$\beta$, although the latter is strongly affected by noise. It seems to be a case similar to PKS~$1441+25$, outlined in the previous section, which implies a FSRQ classification.

To summarize, in the first case the finding of a NLS1 was not possible because of a poor EGRET contours probability, which in turn led to a wrong counterpart, while in the second one, the counterpart is the same as for LAT, but a reliable optical spectrum is still missing to confirm the classification. These few possible detections can be explained by the smaller field of view (FOV) of EGRET\footnote{See here \url{https://fermi.gsfc.nasa.gov/science/instruments/table1-1.html} (accessed on 30 August 2022) for a comparison between EGRET and LAT instruments.} ($\sim$0.5~sr vs. $>$2~sr of LAT), which means that almost pointed observations were required to catch a flare of a NLS1. On the opposite, {\em Fermi}/LAT, with its large FOV and excellent sensitivity, can scan the entire sky every three hours, implying a significant increase of the probability to detect an outburst from Seyfert-type jetted AGN. 

\section{The Twilight Zone}
The manual screening of such a large sample of cosmic sources gave us the opportunity to observe many unusual features of these objects. Among the most interesting cases, there~are:

\begin{enumerate}
\item 4FGL~J$1416.1+1320$~=~PKS~B$1413+135$: the jetted AGN is behind a Seyfert~2 galaxy at $z=0.247$ \cite{READHEAD};
\item 4FGL~J$1615.6+4712$~=~B3~$1614+473$: the SDSS image\footnote{\url{http://skyserver.sdss.org/DR16//en/tools/chart/navi.aspx?ra=243.921721316305&dec=47.1866096751966&scale=0.2} (accessed on 30 August 2022).} shows the object forming something like a circle with other apparently close objects, perhaps an Einstein ring?;
\item 4FGL~J$1647.5+4950$~=~SBS~$1646+499$: this jetted AGN is a Seyfert hosted in a spiral galaxy, where a SNII exploded in 2009 (2009fe, see Figure 15 in \cite{HAKOBYAN});
\item 4FGL~J$1744.0+1935$~=~S3~$1741+19$: it is a triple interacting system \cite{HEIDT};
\item 4FGL~J$2204.3+0438$~=~4C~$+04.77$: originally classified as BL Lac Object, because of small equivalent width emission lines, once the host galaxy continuum is removed, it clearly displays a Seyfert-1 spectrum (see Figure 4 in \cite{VERON});
\item 4FGL~J$2302.8-1841$~=~PKS~$2300-18$: tidal interaction with a close companion, precessing jet \cite{HUNSTEAD};
\end{enumerate}

\section{Final Remarks}
We presented a list of 2980 gamma-ray sources from the Fourth {\em Fermi} LAT point-source catalog, with revised classification and spectroscopic or photometric redshift. The main result is that the gamma-ray emitting jetted AGN zoo is more variegated than previously thought, with emerging populations of Seyfert-type AGN. It is also worth noting that an AGN can change classification with time on human time scales, because of intrinsic changes in the emission mechanisms. 

We would like to stress that users should read the literature thoroughly before using their data and conclusions. A simple cross-match of catalogs at different frequencies is not sufficient. A lot of high-level information (which can be found only in published papers, because it required a human analysis) can be missed, with significant impact on the knowledge about the nature of these cosmic sources. In addition, as we already noted in the Paper I, online databases may not be updated with the most recent findings or occasionally contain plain errors.

\vspace{6pt} 

\section*{Acknowledgments}
We thank Marco Simonte for help with the extraction of photometric redshift values from NOIRlab. Funding for the Sloan Digital Sky Survey (SDSS) was provided by the Alfred P. Sloan Foundation, the Participating Institutions, the National Aeronautics and Space Administration, the National Science Foundation, the U.S. Department of Energy, the Japanese Monbukagakusho, and the Max Planck Society. The SDSS Web site is \url{http://www.sdss.org/}  (accessed on 30 August 2022). The SDSS is managed by the Astrophysical Research Consortium (ARC) for the Participating Institutions. The Participating Institutions are the University of Chicago, Fermilab, the Institute for Advanced Study, the Japan Participation Group, The Johns Hopkins University, Los Alamos National Laboratory, the Max--Planck--Institute for Astronomy (MPIA), the Max--Planck--Institute for Astrophysics (MPA), New Mexico State University, University of Pittsburgh, Princeton University, the United States Naval Observatory, and the University of Washington. Guoshoujing Telescope (the Large Sky Area Multi-Object Fiber Spectroscopic Telescope LAMOST) is a National Major Scientific Project built by the Chinese Academy of Sciences. Funding for the project has been provided by the National Development and Reform Commission. LAMOST is operated and managed by the National Astronomical Observatories, Chinese Academy of Sciences. This research  made use of the SIMBAD database, operated at CDS, Strasbourg, France (2000, A\&AS, 143, 9, ``The SIMBAD astronomical database'', Wenger et al.). This research has made use of the VizieR catalogue access tool, CDS, Strasbourg, France (DOI : 10.26093/cds/vizier). The original description of the VizieR service was published in 2000, A\&AS 143, 23. This research made use of the NASA/IPAC Extragalactic Database (NED), which is funded by the National Aeronautics and Space Administration and operated by the California Institute of Technology. This research  made use of NASA’s Astrophysics Data System Bibliographic Services.


\begin{thebibliography}{999}
\bibitem{REES} Rees, M.J. Appearance of Relativistically Expanding Radio Sources. {\em Nature} {\bf 1966}, {\em 211}, 468--470.
\bibitem{SCHMIDT} Schmidt, M. Space Distribution and Luminosity Functions of Quasi-Stellar Radio Sources. {\em Astrophys. J.} {\bf 1968}, {\em 151}, 393--409.
\bibitem{FR} Fanaroff, B.L.; Riley, J.M. The morphology of extragalactic radio sources of high and low luminosity. {\em Mon. Not. R. Astron. Soc.} {\bf 1974}, {\em 167}, 31P--35P.
\bibitem{BLANDREES} Blandford, R.D.; Rees, M.J. Some comments on radiation mechanisms in Lacertids. In {\em Pittsburgh Conference on BL Lac Objects}; Wolfe, A.M., Ed.; University of Pittsburgh: Pittsburgh, PA, USA, {1978}; pp. 328--347.
\bibitem{ORRBROWNE} Orr, M.J.L.; Browne, I.W.A. Relativistic beaming and quasar statistics. {\em Mon. Not. R. Astron. Soc.} {\bf 1982}, {\em 200}, 1067--1080.
\bibitem{BARTHEL} Barthel, P.D. Is every quasar beamed? {\em  Astrophys. J.} {\bf 1989}, {\em 336}, 606--611.
\bibitem{URRYPAD} Urry, C.M.; Padovani, P. Unified Schemes for Radio-Loud Active Galactic Nuclei. {\em Publ. Astron. Soc. Pac.} {\bf 1995}, {\em 107}, 803--845. 
\bibitem{GHISELLINI98} Ghisellini, G.; Celotti, A.; Fossati, G.; Maraschi, L.; Comastri, A. A theoretical unifying scheme for gamma-ray bright blazars. {\em Mon. Not. R. Astron. Soc.} {\bf 1998}, {\em 301}, 451--468.
\bibitem{PADOVANI} Padovani, P. On the two main classes of active galactic nuclei. {\em Nat. Astron.} {\bf 2017}, {\em 1}, 0194.
\bibitem{BLANDFORD} Blandford, R.; Meier, D.; Readhead, A., Relativistic Jets from Active Galactic Nuclei. {\em Annu. Rev. Astron. Astrophys.} {\bf 2019}, {\em 57}, 467--509.
\bibitem{KOMOSSA18} Komossa, S. Multi-wavelength properties of radio-loud Narrow-line Seyfert 1 galaxies. {\em Proc. Sci.} {\bf 2018}, {\em 328}, 15.
\bibitem{FOSCHINI20} Foschini, L. Jetted Narrow-Line Seyfert 1 Galaxies \& Co.: Where Do We Stand? {\em Universe} {\bf 2020}, {\em 6}, 136.
\bibitem{BERTON16} Berton, M.; Caccianiga, A.; Foschini, L.; Peterson, B.; Mathur, S.; Terreran, G.; Ciroi, S.; Congiu, E.; Cracco, V.; Frezzato, M.; et al. Compact steep-spectrum sources as the parent population of flat-spectrum radio-loud narrow-line Seyfert 1 galaxies. {\em Astron.~Astrophys.} {\bf 2016}, {\em 591}, A98.
\bibitem{FOSCHINI17} Foschini, L.  What we talk about when we talk about blazars? {\em Front. Astron. Space Sci.} {\bf 2017}, {\em 4},  6.
\bibitem{ROMANO18} Romano, P.; Vercellone, S.; Foschini, L.; Tavecchio, F.; Landoni, M.; Kn\"odlseder, J. Prospects for gamma-ray observations of narrow-line Seyfert 1 galaxies with the Cherenkov Telescope Array. {\em Mon. Not. R. Astron. Soc.} {\bf 2018}, {\em 481}, 5046--5061.
\bibitem{4FGL} Abdollahi, S. et al. [Fermi LAT Collaboration]. Fermi Large Area Telescope Fourth Source Catalog. {\em  Astrophys. J. Suppl. Ser.} {\bf 2020}, {\em 247}, 33. 
\bibitem{4LAC} Ajello, M. et al. [Fermi LAT Collaboration]. The Fourth Catalog of Active Galactic Nuclei detected by the Fermi Large Area Telescope. {\em  Astrophys. J.} {\bf 2020}, {\em 892}, 105.
\bibitem{FOSCHINI21} Foschini, L.; Lister, M.L., Ant\'on, S.; Berton, M.; Ciroi, S.; March\~a, M.J.M.; Tornikoski, M.; Järvelä, E.; Romano, P.; Vercellone, S.; Dalla Bont\'a, E. A New Sample of Gamma-Ray Emitting Jetted Active Galactic Nuclei -- Preliminary Results. {\em Universe} {\bf 2021}, {\em 7}, 372.
\bibitem{GHISELLINI} Ghisellini, G.; Tavecchio, F.; Foschini, L.; Bonnoli, G.; Tagliaferri, G. The red blazar PMN J2345-1555 becomes blue. {\em Mon. Not. R. Astron. Soc. Lett.} {\bf 2013}, {\em 432}, L66--L70. 
\bibitem{HERNANDEZ} Hernández-García, L.; Panessa, F.; Giroletti, M.; Ghisellini, G.; Bassani, L.; Masetti, N.;  Pović, M.; Bazzano, A.; Ubertini, P.; Malizia, A.; et al. Restarting activity in the nucleus of PBC~J2333.9-2343. An extreme case of jet realignment. {\em Astron. Astrophys.} {\bf 2017}, {\em 603}, A131.
\bibitem{KAUR19} Kaur, A.; Falcone, A.D.; Stroh, M.D.; Kennea, J.A.; Ferrara, E.C. Classification of New X-Ray Counterparts for Fermi Unassociated Gamma-Ray Sources Using the Swift X-Ray Telescope. {\em Astrophys. J.} {\bf 2019}, {\em 887}, 18. 
\bibitem{KERBY21} Kerby, S.; Kaur, A.; Falcone, A.D.; Stroh, M.C.; Ferrara, E.C.; Kennea, J.A.; Colosimo, J. X-Ray Spectra and Multiwavelength Machine Learning Classification for Likely Counterparts to Fermi 3FGL Unassociated Sources. {\em Astron. J.} {\bf 2021}, {\em 161}, 154.
\bibitem{NORRIS21} Norris, R.P.; Marvil, J.; Collier, J.D.; Kapi\'nska, A.D.; O'Brien, A.N.; Rudnick, L.; Andernach, H.; Asorey, J.; Brown, M.J.I.; Br\"uggen, M.; et al. The Evolutionary Map of the Universe pilot survey. {\em Publ. Astron. Soc. Australia} {\bf 2021}, {\em 38}, e046.
\bibitem{SHAW12} Shaw, M.S.; Romani, R.W.; Cotter, G.; Healey, S.E.; Michelson, P.F.; Readhead, A.C.S.; Richards, J.L.; Max-Moerbeck, W.; King, O.G.; Potter, W.J. Spectroscopy of Broad-line Blazars from 1LAC. {\em Astrophys. J.} {\bf 2012}, {\em 748}, 49.
\bibitem{RAKSHIT20} Rakshit, S.; Stalin, C.S.; Kotilainen, J. Spectral Properties of Quasars from Sloan Digital Sky Survey Data Release 14: The Catalog. {\em Astrophys. J. Suppl. Ser.} {\bf 2020}, {\em 249}, 17.
\bibitem{MAGIC15} Ahnen, M.L.; Ansoldi, S.; Antonelli, L.A.; Antoranz, P.; Babic, A.; Banerjee, B.; Bangale, P.; Barres de Almeida, U.; Barrio, J.A.; \mbox{Bednarek, W.; et al.} (MAGIC Coll.) Very High Energy $\gamma$-Rays from the Universe's Middle Age: Detection of the $z=0.940$ Blazar PKS~$1441+25$ with MAGIC. {\em Astrophys. J. Lett.} {\bf 2015}, {\em 815}, L23.
\bibitem{CHEN89} Chen, K.; Halpern, J.P. Structure of Line-emitting Accretion Disks in Active Galactic Nuclei: ARP 102B. {\em Astrophys. J.} {\bf 1989}, {\em 344}, 115--124.
\bibitem{SHENHO} Shen, Y.; Ho, L.C. The diversity of quasars unified by accretion and orientation. {\em Nature} {\bf 2014}, {\em 513}, 210--213. 
\bibitem{PANDA} Panda, S.; Marziani, P.; Czerny, B. The Quasar Main Sequence Explained by the Combination of Eddington Ratio, Metallicity, and Orientation. {\em Astrophys. J.} {\bf 2019}, {\em 882}, 79.
\bibitem{MARZIANI22} Marziani, P.; Bon, E.; Bon, N.; D'Onofrio, M.; Punsly, B.; \'Sniegowska, M.; Czerny, B.; Panda, S.; Martnez Aldama, M.L.; del Olmo, A.; et al. The main sequence of quasars: The taming of the extremes. {\em Astron. Nach.} {\bf 2022}, {\em 343}, e210082. 
\bibitem{MARZIANI13} Marziani, P.; Sulentic, J.W.; Plauchu-Frayn, I.; del Olmo, A. Low-ionization Outflows in High Eddington Ratio Quasars. {\em Astrophys. J.} {\bf 2013}, {\em 764}, 150.
\bibitem{PUNSLY20} Punsly, B.; Marziani, P.; Berton, M.; Kharb, P. The Extreme Red Excess in Blazar Ultraviolet Broad Emission Lines. {\em Astrophys. J.} {\bf 2020}, {\em 903}, 44.
\bibitem{HO14} Ho, L.C.; Kim, M. The Black Hole Mass Scale of Classical and Pseudo Bulges in Active Galaxies. {\em Astrophys. J.} {\bf 2014}, {\em 789}, 17.
\bibitem{BERTON15} Berton, M.; Foschini, L.; Ciroi, S.; Cracco, V.; La Mura, G.; Lister, M.L.; Mathur, S.; Peterson, B.M.; Richards, J.L.; Rafanelli, P. Parent population of flat-spectrum radio-loud narrow-line Seyfert 1 galaxies. {\em Astron. Astrophys.} {\bf 2015}, {\em 578}, A28.
\bibitem{HEINZ} Heinz, S.; Sunyaev, R.A. The non-linear dependence of flux on black hole mass and accretion rate in core-dominated jets. {\em Mon. Not. R. Astron. Soc.} {\bf 2003}, {\em 343}, L59--L64.
\bibitem{ABDO09} Abdo, A.A. et al. [Fermi LAT Collaboration]. Fermi/Large Area Telescope discovery of gamma-ray emission from a relativistic jet in the narrow-line quasar PMN~J$0948+0022$. {\em  Astrophys. J.} {\bf 2009}, {\em 699}, 976--984.
\bibitem{FOSCHINI11} Foschini, L.; Ghisellini, G.; Kovalev, Y.Y.; Lister, M.L.; D'Ammando, F.; Thompson, D.J.; Tramacere, A.; Angelakis, E.; Donato, D.; Falcone, A.; et al. The first gamma-ray outburst of a narrow-line Seyfert 1 galaxy: the case of PMN J0948+0022 in 2010 July. {\em Mon. Not. R. Astron. Soc.} {\bf 2011}, {\em 413}, 1671--1677.
\bibitem{PALIYA14} Paliya, V.S.; Sahayanathan, S.; Parker, M.L.; Fabian, A.C.; Stalin, C.S.; Anjum, A.; Pandey, S.B. The Peculiar Radio-loud Narrow Line Seyfert 1 Galaxy 1H~$0323+342$. {\em  Astrophys. J.} {\bf 2014}, {\em 789}, 143.
\bibitem{PALIYA16} Paliya, V.S.; Stalin, C.S. The First GeV Outburst of the Radio-loud Narrow-line Seyfert 1 Galaxy PKS~$1502+036$. {\em  Astrophys. J.} {\bf 2016}, {\em 820}, 52.
\bibitem{EGRET} Hartman, R.C.; Bertsch, D.L.; Bloom, S.D.; Chen, A.W.; Deines-Jones, P.; Esposito, J.A.; Fichtel, C.E.; Friedlander, D.P.; Hunter, S.D.; McDonald, L.M.; et al. The Third EGRET Catalog of High-Energy Gamma-Ray Sources. {\em  Astrophys. J. Suppl. Ser.} {\bf 1999}, {\em 123}, 79--202.
\bibitem{WU22} Wu, Q.; Shen, Y. A Catalog of Quasar Properties from Sloan Digital Sky Survey Data Release 16. {\em arXiv} {\bf 2022}, arXiv:2209.03987.
\bibitem{EGRET2} Nandikotkur, G.; Jahoda, K.M.; Hartman, R.C.; Mukherjee, R.; Sreekumar, P.; B\"ottcher, M.; Sambruna, R.M.; Swank, J.H. Does the Blazar Gamma-Ray Spectrum Harden with Increasing Flux? Analysis of 9 Years of EGRET Data. {\em Astrophys. J.} {\bf 2007}, {\em 657}, 706--724.
\bibitem{READHEAD} Readhead, A.C.S.; Ravi, V.; Liodakis, I.; Lister, M.L.; Singh, V.; Aller, M.F.; Blandford, R.D.; Browne, I.W.A.; Gorjian, V.; Grainge, K.J.B.; et al. The Relativistic Jet Orientation and Host Galaxy of the Peculiar Blazar PKS~$1413+135$. {\em Astrophys. J.} {\bf 2021}, {\em 907}, 61. 
\bibitem{HAKOBYAN} Hakobyan, A.A.; Adibekyan, V.Z.; Aramyan, L.S.; Petrosian, A.R.; Gomes, J.M.; Mamon, G.A.; Kunth, D.; Turatto, M. Supernovae and their host galaxies. I. The SDSS DR8 database and statistics. {\em Astron. Astrophys.} {\bf 2012}, {\em 544}, A81.
\bibitem{HEIDT} Heidt, J.; Nilsson, K.; Fried, J.W.; Takalo, L.O.; Sillanp\"a\"a, A. 1ES 1741+196: a BL Lacertae object in a triplet of interacting galaxies? {\em Astron. Astrophys.} {\bf 1999}, {\em 348}, 113--116. 
\bibitem{VERON} V\'eron-Cetty, M.-P.; V\'eron, P. Spectroscopic observatioons of sixteen BL Lacertae candidates. {\em Astron. Astrophys. Suppl. Ser.} {\bf 1993}, {\em 100}, 521--529.
\bibitem{HUNSTEAD} Hunstead, R.W.; Murdoch, H.S.; Condon, J.J.; Phillips, M.M. A QSO with precessing jets: $2300-189$. {\em Mon. Not. R. Astron. Soc.} {\bf 1984}, {\em 207}, 55--71. 
\end{thebibliography}
\end{document}